\documentclass{article}
\usepackage[T1]{fontenc}
\usepackage{bookman}
\usepackage{geometry}
\usepackage{graphics}

\makeatletter

\providecommand{\LyX}{L\kern-.1667em\lower.25em\hbox{Y}\kern-.125emX\@}

\usepackage{a4}
\input{epsfig.sty}
\def\fnum@table{\tablename~{\bf\thetable}}
\def\fnum@figure{\figurename~{\bf\thefigure}}
\def\tablename{\footnotesize{\bf Table}}
\def\figurename{\footnotesize{\bf Figure}}

\def\be{\begin{equation}}
\def\ee{\end{equation}}

\makeatother

\begin{document}

\title{\textbf{\huge Initial Condition for QGP Evolution }\\
\textbf{\huge from N}\textbf{\textsc{\huge E}}\textbf{\textsc{\Huge X}}\textbf{\textsc{\huge US}}
\textbf{\textsc{\huge }}\\
\textbf{\textsc{}}\\
}

\author{\textbf{H.J. Drescher\protect\( ^{1,4}\protect \), F.M. Liu}\protect\( ^{1,5}\protect \)\textbf{,
S. Ostapchenko\protect\( ^{1,2,3}\protect \), T. Pierog\protect\( ^{1}\protect \),
and K. Werner}\protect\( ^{1}\protect \)\\
\\
\\
 \textit{\protect\( ^{1}\protect \)} \textit{\small SUBATECH, Université de
Nantes -- IN2P3/CNRS -- EMN,  Nantes, France }\\
\textit{\small \protect\( ^{2}\protect \) Moscow State University, Institute
of Nuclear Physics, Moscow, Russia}\\
 \textit{\small \protect\( ^{3}\protect \) Institut f. Kernphysik, Forschungszentrum
Karlsruhe, Karlsruhe, Germany}\\
\textit{\small \protect\( ^{4}\protect \) Physics Department, New York University,
New York, USA} \textit{ }\\
\textit{\small \protect\( ^{5}\protect \) Institute of Particle Physics, Huazhong
Normal University, Wuhan, China}\textit{}\\
}

\date{\vspace{-0.5cm }}

\maketitle
\begin{abstract}
We recently proposed a new approach to high energy nuclear scattering, which
treats the initial stage of heavy ion collisions in a sophisticated way. We
are able to calculate macroscopic quantities like energy density and velocity
flow at the end of this initial stage, after the two nuclei having penetrated
each other. In other words, we provide the initial conditions for a macroscopic
treatment of the second stage of the collision. We address in particular the
question of how to incorporate the soft component properly. We find almost perfect
``Bjorken scaling'': the rapidity coincides with the space-time rapidity,
whereas the transverse flow is practically zero. The distribution of the energy
density in the transverse plane shows typically a very ``bumpy'' structure.
\end{abstract}

\section{Introduction}

Unfortunately there does not exist a single formalism able to account for a
complete nucleus-nucleus collision. Rather we have to -- at least for the moment
-- divide the reaction into different stages
\begin{figure}[htb]
{\par\centering \vspace{-1cm}\par \resizebox*{!}{0.35\textheight}{\rotatebox{270}{\includegraphics{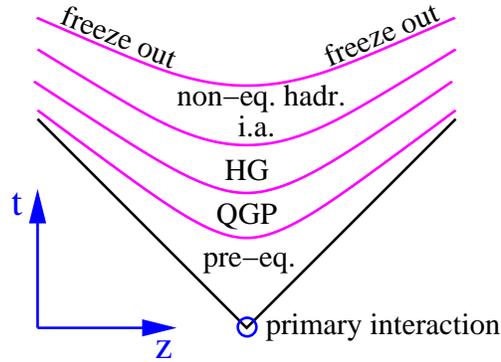}}} \\
\vspace{-1cm}\par\par}

\caption{The different stages of heavy ion collisions.\label{stages}}
\end{figure}
(see fig.\ \ref{stages}) and try to understand them \textbf{\textit{}}as well
as possible. There is, first of all, the primary interaction when the two nuclei
pass through each other. Since at very high energies the longitudinal size is,
due to the gamma factor, almost zero (of the order 0.1 fm at RHIC), all the
nucleons of the projectile interact with all the nucleons of the target instantaneously.
In such a primary interaction many partons are created, which interact (in the
pre-equilibrium stage) before reaching an equilibrium, referred to as quark-gluon
plasma. The system then expands, passing via phase transition (or sudden crossover)
into the hadron gas stage. The density decreases further till the collision
rate is no longer large enough to maintain chemical equilibrium, but there are
still hadronic interactions till finally the particles ``freeze out'', i.e.
they continue their way without further interactions.

The equilibrium stage is often treated macroscopically, by solving hydrodynamical
equations. Here, usually very simplified initial conditions are used, like flat
distributions in space-time rapidity. This could be done much better by calculating
the initial conditions for the hydrodynamical evolution on the basis of a realistic
model for the primary interactions.

We recently presented a completely new approach \cite{wer97,dre99a,dre00} for
hadronic interactions and the initial stage of nuclear collisions, referred
to as \textsc{neXus}, where we provide a rigorous treatment of the multiple
scattering aspect. Questions of energy conservation are clearly determined by
the rules of field theory, both for cross section and particle production calculations,
which is not the case in all of the corresponding models used so far \textbf{\textit{}}to
calculate the initial stage. In addition, we introduced (currently only to leading
order) so-called enhanced diagrams, responsible for screening and diffraction.
It was not the idea to create another model with some more features, but to
provide a model which is \textbf{\textit{}}theoretically \textbf{\textit{}}consistent,
and therefore much more realistic than all the approaches used before. We are
therefore using \textsc{neXus} in order to determine macroscopic quantities
after the first stage, when the two nuclei have traversed each other, such that
these quantities may be used as initial conditions for \textbf{\textit{}}a macroscopic
treatment of the later stages of the collision.

\begin{figure}[htb]
{\par\centering \resizebox*{!}{0.25\textheight}{\includegraphics{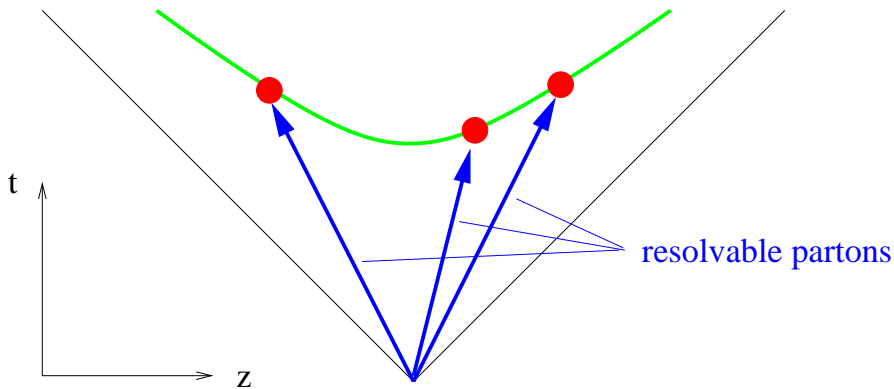}} \par}

\caption{The naive method: only the resolvable partons are considered to determine the
energy density on a hyper-surface (dots on hyperbola).\label{fig:dens_wrong}}
\end{figure}

\begin{figure}[htb]
{\par\centering \resizebox*{!}{0.25\textheight}{\includegraphics{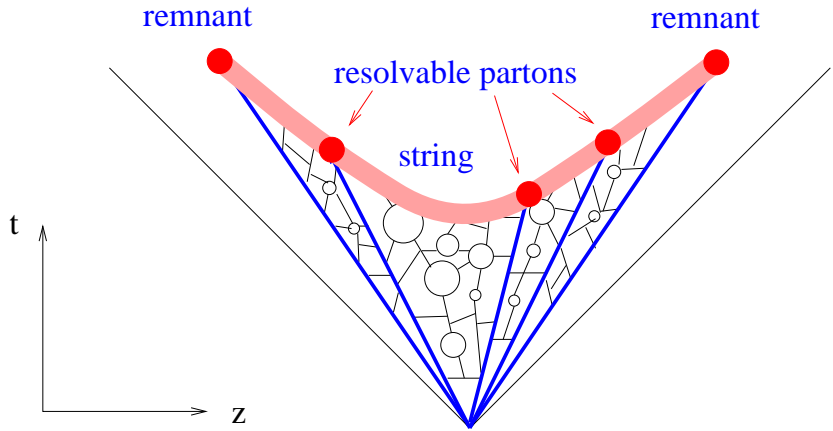}} \par}

\caption{Between resolvable partons (thick lines) there are in addition unresolvable
partons (thin lines), which are not treated explicitly. Their \textbf{\textit{}}contribution
is considered implicitly via strings.\label{fig:dens_right}}
\end{figure}

Calculating for example energy densities from a model like \textsc{neXus} or
any other model for primary interactions, is not trivial. We know the momenta
of all the partons and we may calculate the energy density of the partonic system
on a given hyper-surface (constant \( \tau  \)), as shown in fig.\ \ref{fig:dens_wrong}. 
Here only the ``resolvable'' (or hard) partons are considered (dots on the
hyperbola representing a hyper-surface). But this is certainly not the correct
answer for the quantity of interest, since there are many ``unresolved'' (or
soft) partons around, which contribute to the energy density in a significant
fashion. The problem can be solved by using the string model, which is nothing
but an attempt to treat the ``soft partons'' implicitly . The soft partons
represent the string between ``kinks'', the latter ones representing the hard
partons, see fig.\ \ref{fig:dens_right}. 
Thus we are able to calculate energy densities in nucleus-nucleus collisions
taking properly into account soft and hard (resolvable and unresolvable) contributions.

The outline of this paper is as follows: we first review briefly the basic elements
of the \textsc{neXus} model, before we discuss in somewhat more detail the role
of strings in this approach and the dynamics of strings. We then proceed to
calculate energy densities for a given proper time, taking into account the
hard and the soft partons.

\section{The NE{\LARGE X}US Model \label{NM label}}

The most sophisticated approach to high energy hadronic interactions is the
so-called Gribov-Regge theory \textbf{}\cite{gri68}. This is an effective field
theory, which allows multiple interactions to happen ``in parallel'', with
phenomenological objects called ``Pomerons'' representing elementary interactions
\cite{bak76}. Using the general rules of field theory, one may express cross
sections in terms of a couple of parameters characterizing the Pomeron. Interference
terms are crucial, as they assure the unitarity of the theory. 

A big disadvantage is the fact that cross sections and particle production are
not calculated consistently: the fact that energy needs to be shared between
many Pomerons in case of multiple scattering is well taken into account when
considering particle production (in particular in Monte Carlo applications),
but not for cross sections \cite{abr92}. 

Another problem is the fact that at high energies, one also needs a consistent
approach to include both soft and hard processes. The latter ones are usually
treated in the framework of the parton model, which only allows the calculation
of inclusive cross sections.

We recently presented a completely new approach \cite{wer97,dre99a,dre00} for
hadronic interactions and the initial stage of nuclear collisions, which is
able to solve several of the above-mentioned problems. We provide a rigorous
treatment of the multiple scattering aspect, such that questions of energy conservation
are clearly determined by the rules of field theory, both for cross section
and particle production calculations. In both (!) cases, energy is properly
shared between the different interactions happening in parallel. This is the
most important new aspect of our approach, which we consider a first necessary
step to construct a consistent model for high energy nuclear scattering. 

\begin{figure}[htb]
{\par\centering \resizebox*{!}{0.11\textheight}{\includegraphics{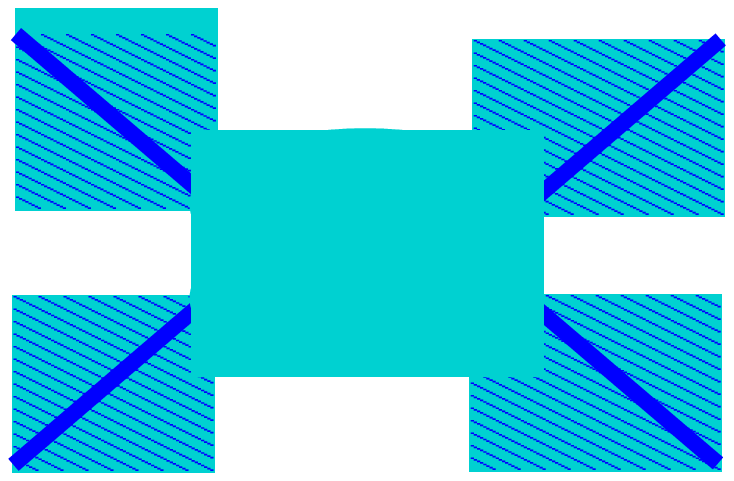}} \resizebox*{!}{0.15\textheight}{\includegraphics{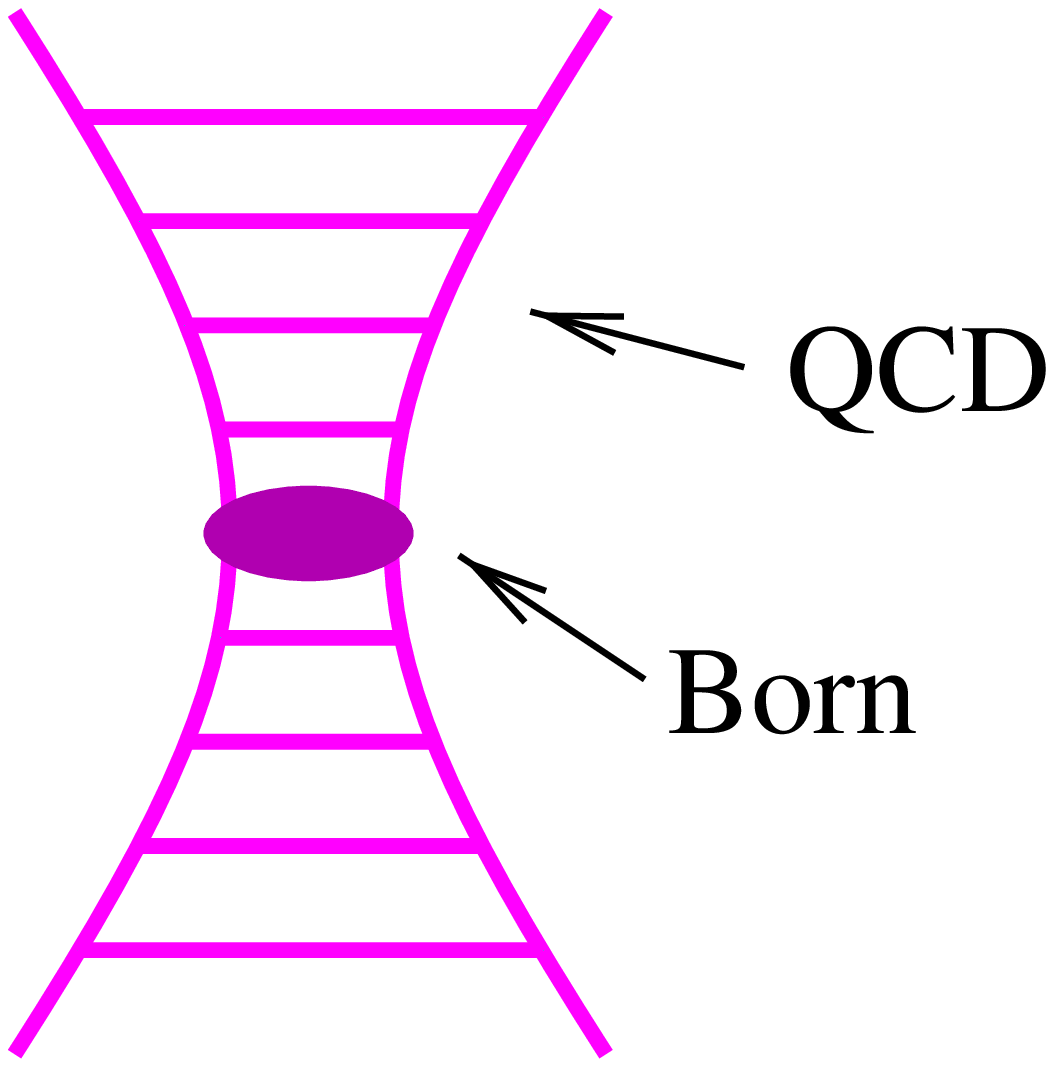}} \resizebox*{!}{0.15\textheight}{\includegraphics{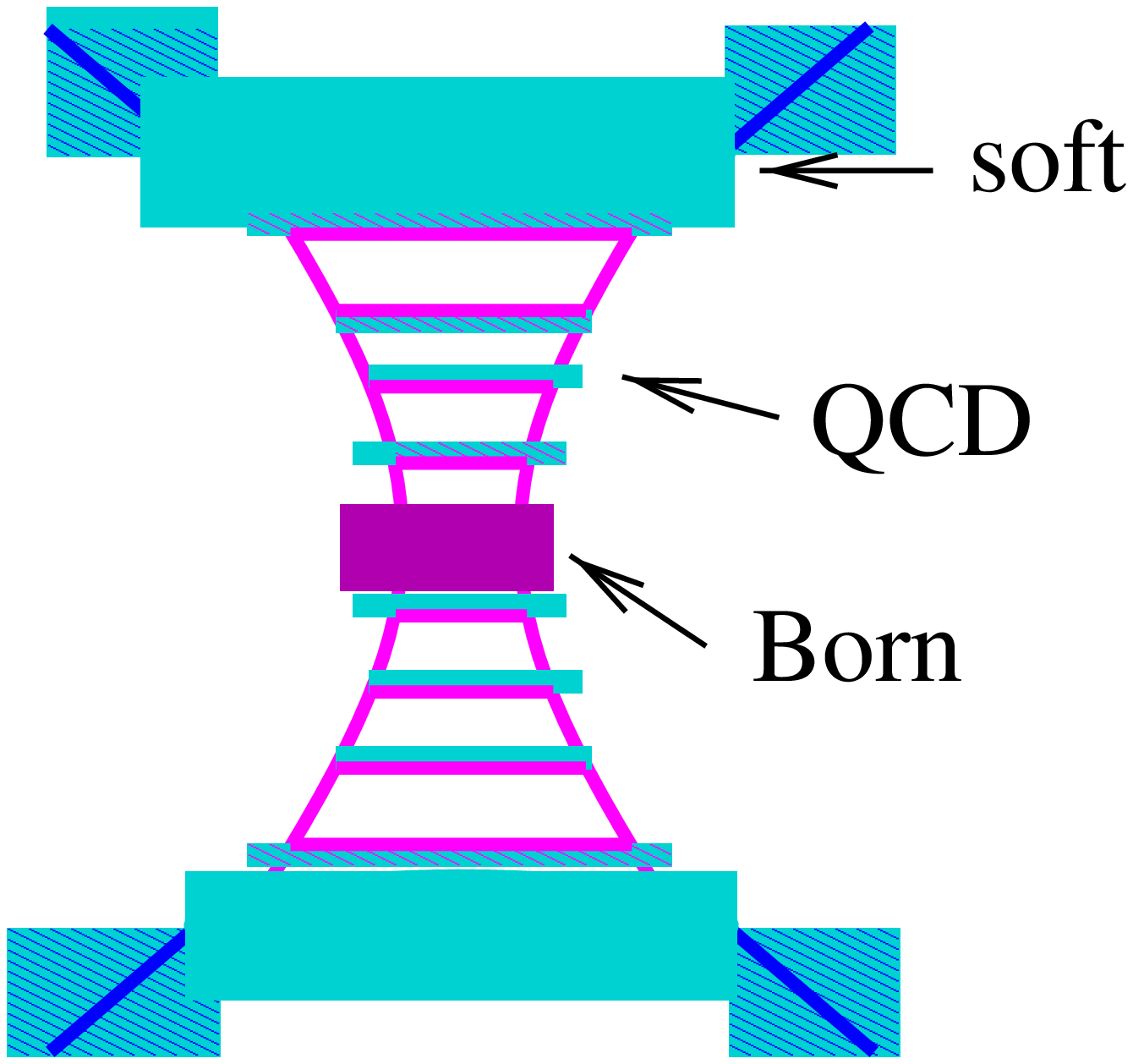}} \par}

\caption{The soft elastic scattering amplitude \protect\( T_{\mathrm{soft}}\protect \)
(left), the hard elastic scattering amplitude \protect\( T_{\mathrm{hard}}\protect \)
(middle) and one of the three contributions to the semi-hard elastic scattering
amplitude \protect\( T_{\mathrm{semi}}\protect \) (right). \label{nn}}
\end{figure}

We first consider \( pp \) scattering. An elementary interaction is given as
a sum of soft, semi-hard, and hard contributions: \( T_{2\rightarrow 2}=T_{\mathrm{soft}}+T_{\mathrm{semi}}+T_{\mathrm{hard}} \),
as discussed in detail in ref. \cite{dre00}. We have a hard contribution \( T_{\mathrm{hard}} \),
when the the first partons on both sides are valence quarks, a semi-hard contribution
\( T_{\mathrm{semi}} \), when at least on one side there is a sea quark (being
emitted from a soft Pomeron), and finally we have a soft contribution, when
there is no hard scattering at all (see fig. \ref{nn}). 
\( T_{\mathrm{hard}} \) is calculated using the standard techniques of perturbative
QCD, \( T_{\mathrm{soft}} \) is parameterized, and \( T_{\mathrm{semi}} \)
is calculated as a convolution of \( T_{\mathrm{soft}} \) and \( T_{\mathrm{hard}} \).
We have a smooth transition from soft to hard physics: at low energies the soft
contribution dominates, at high energies the hard and semi-hard ones, at intermediate
energies (that is where experiments are performed presently) all contributions
are important.

Let us consider nucleus-nucleus (\( AB \)) scattering. The nucleus-nucleus
scattering amplitude is defined by the sum of contributions of diagrams, corresponding
to multiple elementary scattering processes between parton constituents of projectile
and target nucleons. These elementary scatterings are the sum of soft, semi-hard,
and hard contributions: \( T_{2\rightarrow 2}=T_{\mathrm{soft}}+T_{\mathrm{semi}}+T_{\mathrm{hard}} \).
A corresponding relation holds for the inelastic amplitude \( T_{2\rightarrow X} \).
We introduce ``cut elementary diagrams'' as being the sum over squared inelastic
amplitudes, \( \sum _{X}(T_{2\rightarrow X}) \)\( (T_{2\rightarrow X})^{*} \),
which are graphically represented by vertical dashed lines, whereas the elastic
amplitudes are represented by unbroken lines:

\vspace{0.3cm}
{\par\centering \resizebox*{!}{0.05\textheight}{\includegraphics{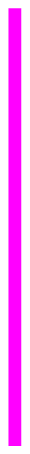}} \( \quad  \)\( =T_{2\rightarrow 2} \),
\( \quad  \)\( \quad  \)\resizebox*{!}{0.05\textheight}{\includegraphics{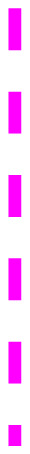}} \( \quad  \)\( =\sum _{X}(T_{2\rightarrow X}) \)\( (T_{2\rightarrow X})^{*} \).\par}
\vspace{0.3cm}

\begin{figure}[htb]
{\par\centering \resizebox*{!}{0.2\textheight}{\includegraphics{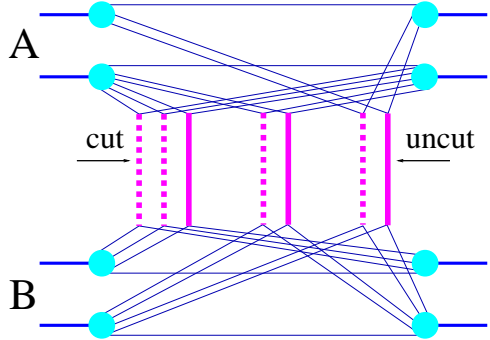}} \par}

\caption{Example of a cut multiple scattering diagram, with cut (dashed lines) and uncut
(full lines) elementary diagrams (Pomerons). \label{grtppaac}}
\end{figure}

\noindent This is very handy for treating the nuclear scattering model. We define
the model via the elastic scattering amplitude \( T_{AB\rightarrow AB} \) which
is assumed to consist of purely parallel elementary interactions between partonic
constituents, described by \( T_{2\rightarrow 2} \). The amplitude is therefore
a sum of many terms. Having defined elastic scattering, inelastic scattering
and particle production is practically given, if one employs a quantum mechanically
self-consistent picture. Let us now consider inelastic scattering: one has of
course the same parallel structure, just some of the elementary interactions
may be inelastic, some elastic. The inelastic amplitude being a sum over many
terms -- \( T_{AB\rightarrow X}=\sum _{i}T^{(i)}_{AB\rightarrow X} \) -- has
to be squared and summed over final states in order to get the inelastic cross
section, which provides interference terms \( \sum _{X}(T^{(i)}_{AB\rightarrow X})(T^{(j)}_{AB\rightarrow X})^{*} \).
These can be conveniently expressed in terms of the cut and uncut elementary
diagrams, as shown in fig. \ref{grtppaac}. 
 One has to be careful about energy conservation: all the partonic constituents
(lines) leaving a nucleon (blob) have to share the momentum of the nucleon.
So, in the explicit formula one has an integration over momentum fractions of
the partons, taking care of momentum conservation. This formula is the master
formula of the approach, allowing calculations of cross sections as well as
particle production. In the latter case, the master formula provides probability
distributions for the momenta taken by the Pomerons and the remnants. A very
detailed description with many applications and comparisons with data can be
found in \cite{dre00}.

So far we described only the basic version of the model. In reality we also
consider triple Pomeron vertices to lowest order, as discussed in detail in
ref. \cite{dre00}. We do not yet consider higher orders, nor do we consider
the case where the two legs of the triple Pomeron are connected to different
nuclei. All this is work in progress.

\section{Hadronic Structure of Cut Pomerons}

In order to develop our multiple scattering theory, we use a simple graphical
representation of a cut Pomeron, namely a thick vertical dashed line connecting
the external legs representing nucleon components, 
\begin{figure}[htb]
{\par\centering \resizebox*{!}{0.1\textheight}{\includegraphics{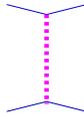}} \par}

\caption{Symbol representing a cut Pomeron.\label{pom}}
\end{figure}
as shown in fig.\ \ref{pom}. This simple diagram hides somewhat the fact that
there is a complicated structure hidden in this Pomeron, and the purpose of
this section is to discuss in particular the internal structure of the Pomeron.

\begin{figure}[htb]
{\par\centering {\huge \resizebox*{!}{0.17\textheight}{\includegraphics{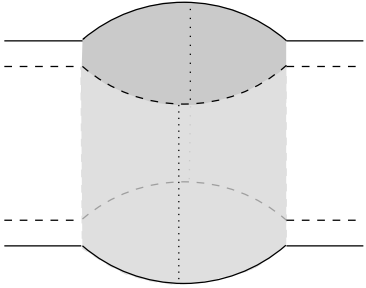}} }\huge \par}

\caption{Cut soft Pomeron represented as a cut cylinder. The grey areas represent unresolved
partons.\label{cyl}}
\end{figure}

Let us start our discussion with the soft Pomeron. Based on Veneziano's topological
expansion one may consider a soft Pomeron as a ``cylinder'', i.e. the sum
of all possible QCD diagrams having a cylindrical topology, see fig.\ \ref{cyl}.
As discussed in detail in \cite{dre00}, the ``nucleon components'' mentioned
earlier, representing the external legs of the diagram, are always quark-anti-quark
pairs, indicated by a dashed line (anti-quark) and a full line (quark) in fig.\
\ref{cyl}. Important for the discussion of particle production are of course
cut diagrams, therefore we show in fig.\ \ref{cyl} a cut cylinder representing
a cut Pomeron: the cut plane is shown as two vertical dotted lines. Let us consider
the half-cylinder, for example, the one to the left of the cut, representing
an inelastic amplitude. 
\begin{figure}[htb]
{\par\centering \resizebox*{!}{0.2\textheight}{\includegraphics{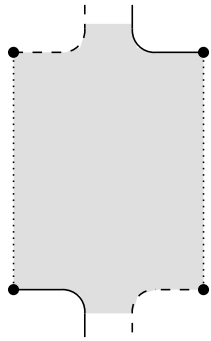}} \par}

\caption{Planar representation of a half-cylinder obtained from cutting a cylinder diagram
(see fig.\ \ref{cyl}). \label{half-cyl}}
\end{figure}
 
\begin{figure}[htb]
{\par\centering \resizebox*{!}{0.2\textheight}{\includegraphics{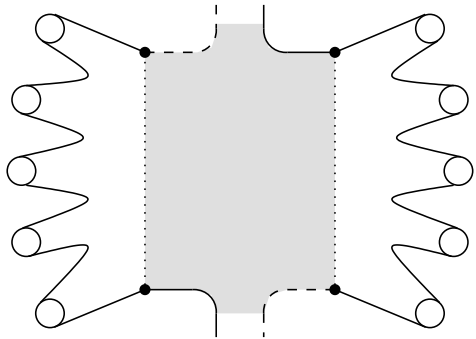}} \par}

\caption{The string model: each cut line (dotted vertical lines) represents a string,
which decays into final state hadrons (circles).\label{string-model}}
\end{figure}
We may unfold this object in order to have a planar representation, as shown
in fig.\ \ref{half-cyl}. Here, the dotted vertical lines indicate the cuts
of the previous figure, and it is there where the hadronic final state hadrons
appear. Lacking a theoretical understanding of this hadronic structure, we simply
apply a phenomenological procedure, essentially a parameterization. We require
the method to be as simple as possible, with a minimum of necessary parameters.
A solution coming close to these demands is the so-called string model: each
cut line is identified with a classical relativistic string. A Lorentz invariant
string breaking procedure provides the transformation into a hadronic final
state, see fig.\ \ref{string-model}.

The phenomenological microscopic picture which stays behind this procedure was
discussed in a number of reviews \cite{kai82,cap94,and83}: the string end-point
partons resulted from the interaction appear to be connected by a color field.
With the partons flying apart, this color field is stretched into a tube, which
finally breaks up giving rise to the production of hadrons and to the neutralization
of the color field.

\begin{figure}[htb]
{\par\centering \resizebox*{!}{0.2\textheight}{\includegraphics{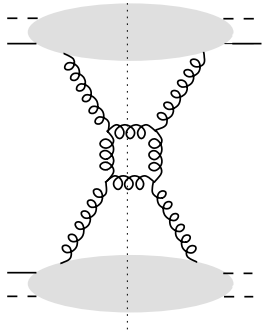}} \par}

\caption{A simple diagram contributing to the semi-hard Pomeron of the ``sea-sea''
type. \label{sea-sea}}
\end{figure}

\begin{figure}[htb]
{\par\centering \resizebox*{!}{0.2\textheight}{\includegraphics{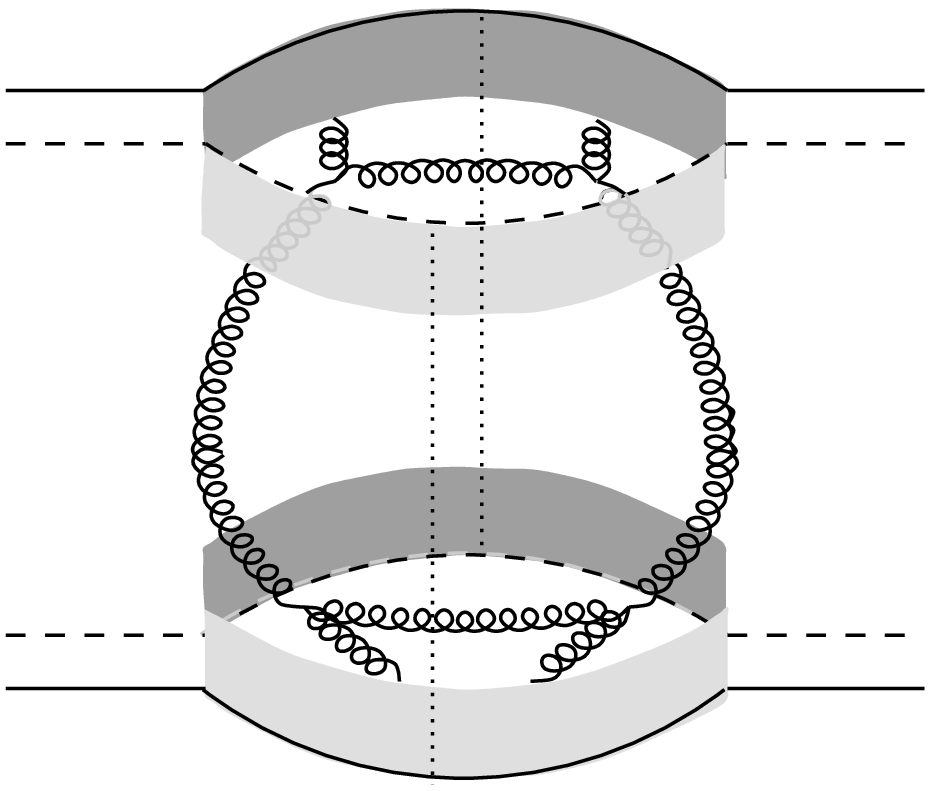}} \( \qquad  \)\( \qquad  \)\resizebox*{!}{0.2\textheight}{\includegraphics{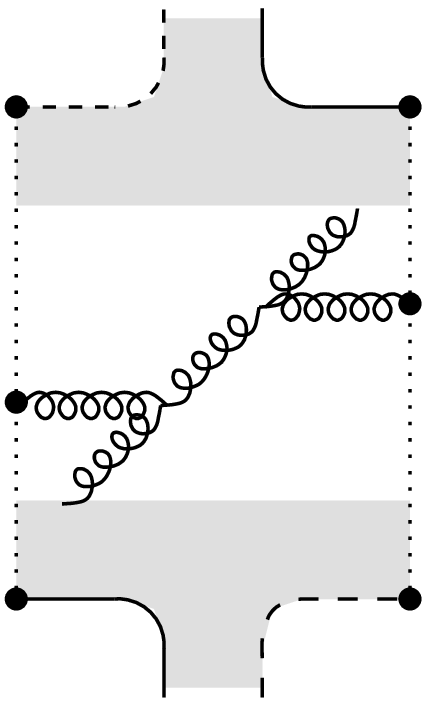}} \par}

\caption{Cylindrical representation of a contribution to the semi-hard Pomeron (left
figure) and planar diagram representing the corresponding half-cylinder (right
figure).\label{sea-sea-cyl}}
\end{figure}

\begin{figure}[htb]
{\par\centering \resizebox*{!}{0.2\textheight}{\includegraphics{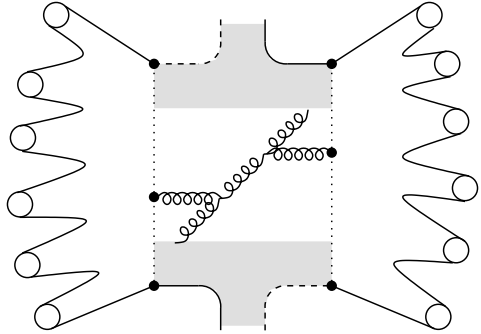}} \par}

\caption{The ``kinky'' string model: the cut line (vertical dotted line) corresponds
to a kinky string, which decays into hadrons (circles). \label{kinky-string-model}}
\end{figure}

We now consider a semi-hard Pomeron of the ``sea-sea'' type, where we have
a hard pQCD process in the middle and a soft evolution at the end, see fig.\ \ref{sea-sea}.
 We generalize the picture introduced above for the soft Pomeron. Again, we
assume a cylindrical structure. For the example of fig.\ \ref{sea-sea}, we
have the picture shown in fig.\ \ref{sea-sea-cyl}: the shaded areas on the
cylinder ends represent the soft Pomerons, whereas in the middle part we draw
explicitly the gluon lines on the cylinder surface. We apply the same procedure
as for the soft Pomeron: we cut the diagram and present a half-cylinder in a
planar fashion, see fig.\ \ref{sea-sea-cyl}.
We observe one difference compared to the soft case: there are three partons
(dots) on each cut line: apart from the quark and the anti-quark at the end,
we have a gluon in the middle. We again apply the string picture, but here we
identify a cut line with a so-called kinky string, where the internal gluons
correspond to internal kinks. The underlying microscopic picture will be presented
by three color-connected partons - the gluon connected by the color field to
the quark and to the anti-quark. The string model provides then a ``parameterization''
of hadron production, see fig.\ \ref{kinky-string-model}.
The procedure described above can be easily generalized to the case of complicated
parton ladders involving many gluons and quark-anti-quark pairs. One should
note that the treatment of semi-hard Pomerons is just a straightforward generalization
of the string model for soft Pomerons, or one might see it the other way round:
the soft string model is a natural limiting case of the kinky string procedure
for semi-hard Pomerons. In a similar way one may treat Pomerons of valence type.

The general picture should be clear from the above examples: in any case, no
matter what type of Pomeron, unresolved soft partons play a very important role.
In the string model, they represent the string pieces between the hard partons.
In case of single Pomeron exchange in proton-proton scattering, particle production
can be treated in a phenomenological fashion via the hadronization of two (in
general kinky) strings. In nuclear collisions, the situation is more complicated,
since we have many Pomerons and consequently many strings -- closely packed
-- which interact with each other. Nevertheless, we can use the string picture
to calculate energy densities, by considering the strings at an earlier stage,
before they hadronize.

It should be noted finally that particle production from cut Pomerons is not
the whole story. Cutting the complete diagram, one has as well to cut the projectile
and target remnants, which may or not be excited. High mass excitations are
as well considered as strings.

\section{Dynamics of Strings}

The string dynamics is derived from the Nambu-Goto Lagrangian, which has been
constructed based on invariance arguments. The corresponding equation of motion
for the string is a wave equation, with a solution \cite{nam69, reb74, sch75, dre00}
\begin{equation}
\label{string-solution}
X(r,t)=X_{0}+\frac{1}{2}\left[ \int _{r-t}^{r+t}g\left( \xi \right) d\xi \right] ,
\end{equation}
for the four-vector \( X(r,t) \), having already assumed that the initial spatial
extension of the string is zero. The quantity \( X_{0} \) represents the formation
point of the string, which coincides with the position of the nucleon-nucleon
interaction being at the origin of the string formation. The space-like variable
\( r \) represents \textbf{\textit{}}the position along the string for given
time \( t \), whereas the function \( g \) defines the initial velocity,
\begin{equation}
g(r)=\dot{X}(r,t)|_{t=0}.
\end{equation}
We will consider here a special class of strings, namely those with a piecewise
constant function \( g \), 
\begin{equation}
g(r)=v_{k}\quad \; \mathrm{for}\: \frac{E_{k-1}}{\kappa }\leq r\leq \frac{E_{k}}{\kappa },\quad \; 1\leq k\leq n
\end{equation}
 for some integer \( n \). We use for the string tension \( \kappa =1 \) GeV/fm.
The set \( \{E_{k}\} \) is a partition of the interval \( [0,E] \), with \( E \)
being the string energy,
\begin{equation}
0=E_{0}<E_{1}<...<E_{n-1}<E_{n}=E,
\end{equation}
 and \( \{v_{k}\} \) represents \( n \) constant 4-vectors. Such strings are
called kinky strings, with \( n \) being the number of kinks, and the \( n \)
vectors \( v_{k} \) being called kink velocities. The function \( g \) must
be symmetric and periodic, with the period \( 2E/\kappa  \). This defines \( g \)
everywhere, and eq. (\ref{string-solution}) is the complete solution of the
string equation, expressed in terms of the initial condition \( g \). In the
case of kinky strings the latter is expressed in terms of the kink velocities
\( \{v_{k}\} \) and the energy partition \( \{E_{k}\} \). 

What has all this to do with cut Pomerons? So far nothing, and to establish
a link, we have to provide some mapping from the language of Pomerons and partons
into the language of strings. We discussed earlier that a cut Pomeron may be
identified with two sequences of partons of the type
\begin{equation}
q-g-g-...-g-\bar{q},
\end{equation}
representing all the partons on a cut line. We identify such a sequence with
a kinky string, by requiring 
\begin{equation}
\mathrm{parton}=\mathrm{kink},
\end{equation}
which means we identify the partons of the above sequence with the kinks of
a kinky string, such that the partition of the energy is given by the parton
energies,
\begin{equation}
E_{k}=\mathrm{energy}\, \mathrm{of}\, \mathrm{parton}\, k
\end{equation}
and the kink velocities are just the parton velocities,
\begin{equation}
v_{k}=\frac{\mathrm{momentum}\, \mathrm{of}\, \mathrm{parton}\, k}{E_{k}}.
\end{equation}
We consider massless partons, so that the energy is equal to the absolute value
of the parton momentum.
\begin{figure}[htb]
{\par\centering \resizebox*{!}{10cm}{\includegraphics{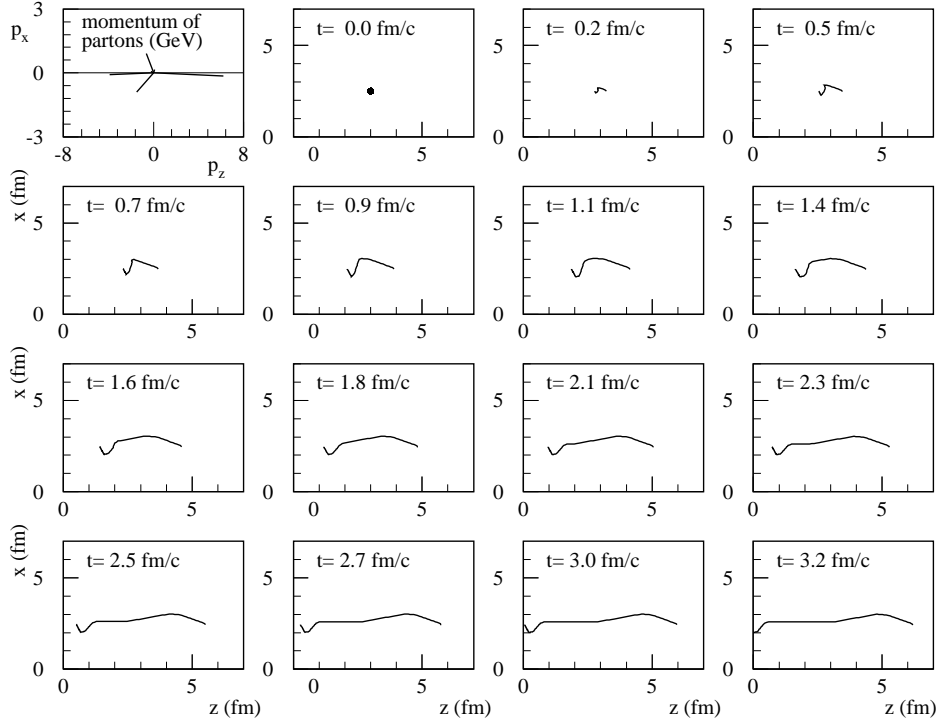}} \par}

\caption{Partons of an \protect\( e^{+}e^{-}\protect \) annihilation event with \protect\( \sqrt{s}=14\textrm{ GeV}\protect \)
in the \protect\( z-x\protect \) plane. The figure in the left upper corner
\textbf{\textit{}}shows the momenta in the \protect\( p_{z}-p_{x}\protect \)
plane\label{fig:kinks_ee}. }
\end{figure}
An example is shown on fig. \ref{fig:kinks_ee}, we have 6 partons -- a quark
and an anti-quark, with 4 gluons in between -- symbolically displayed in the
first sub-figure, with a total cms energy of 14 GeV. One sees that the perturbative
gluons play an important role in the beginning of the movement, and later from
2 GeV on, the longitudinal character dominates. A string breaks typically after
\( 1\textrm{ GeV}/\kappa  \), with \( \kappa  \) being the string tension,
which gives much importance to the perturbative gluons.

\section{Energy Densities from Strings}

Based on the string formalism, we are now able to calculate energy densities.
As discussed above, a string \( S \) is a two-dimensional surface \( X(r,t) \)
in Minkowski space, whose intersection with the surface 
\begin{equation}
t^{2}-z^{2}=\tau ^{2}
\end{equation}
defines uniquely a curve \( S_{\tau } \), which we refer to as ``String at
given proper time \( \tau  \)''. Our aim is to calculate the energy density
and the velocity flow at a given proper time \( \tau  \), in other words, along
the curve \( S_{\tau } \). 

Let us consider a small string piece \( \Delta S_{\tau } \), in other words
a segment of the string curve \( S_{\tau } \) between two neighboring points
\( A \) and \( B \), being sufficiently close so that the string piece may
be considered as point-like. According to relativistic string theory, the four-momentum
of this string piece in the lab system is given as

\begin{equation}
\label{dp}
\Delta P_{\mathrm{lab}}=\int _{A}^{B}\left\{ \frac{\partial X(r,t)}{\partial t}dr+\frac{\partial X(r,t)}{\partial r}dt\right\} .
\end{equation}
 The partial derivatives can be expressed in terms of the initial velocity \( g \)
as
\begin{eqnarray}
\frac{\partial X(r,t)}{\partial t} & = & \frac{1}{2}\left[ g(r+t)+g(r-t)\right] \label{for:xprime} \\
\frac{\partial X(r,t)}{\partial r} & = & \frac{1}{2}\left[ g(r+t)-g(r-t)\right] \, \, .\label{for:xdot} 
\end{eqnarray}
 This velocity function is known, in fact it is defined via the mapping of a
system of \( n \) partons into string language:
\begin{equation}
g(r)=v_{k}\quad \; \mathrm{for}\: \frac{E_{k-1}}{\kappa }\leq r\leq \frac{E_{k}}{\kappa },\quad \; 1\leq k\leq n,
\end{equation}
 where \( v_{k} \) is the four-velocity of the \( k^{\mathrm{th}} \) parton
divided by \( \gamma  \), and the difference \( E_{k}-E_{k-1} \) is its energy.
So the four-momentum of a string segment can be easily expressed in terms of
the original parton momenta. 

We use hyperbolic coordinates,
\begin{equation}
q_{0}=\tau =\sqrt{t^{2}-z^{2}},\qquad q_{1}=x,\qquad q_{2}=y,\qquad q_{3}=\eta =\frac{1}{2}\log \frac{1+z/t}{1-z/t},
\end{equation}
where \( z \) is considered to be the coordinate along the beam axis, and \( t \)
is the time. The variables \( x \) and \( y \) are the transverse coordinates.
For given \( \tau  \) and \( \eta  \), we define a frame \( F_{\eta } \)
via a Lorentz boost with boost rapidity \( \eta  \). The four-momentum of the
above-mentioned string piece is given as 
\begin{equation}
\Delta P^{\mu }=\Lambda _{\nu }^{\mu }\, \Delta P_{\mathrm{lab}}^{\nu }\, ,
\end{equation}
with the corresponding transformation tensor \( \Lambda  \).

We are now going to calculate the energy momentum tensor \( T^{\mu \nu } \)
for a string piece in the frame \( F_{\eta } \). The general definition in
kinetic theory is
\begin{equation}
T^{\mu \nu }(\vec{q})=\int \frac{d^{3}p}{E}p^{\mu }p^{\nu }f(\vec{q},\vec{p}),
\end{equation}
where \( \vec{q} \) is a position four-vector, and \( f \) the phase space
density (particles per phase space volume) for a given time. For our point-like
string piece we have in principle 
\begin{equation}
f(\vec{q},\vec{p})=\delta (\vec{p}-\overrightarrow{\Delta P})\delta (\vec{q}-\vec{Q}).
\end{equation}
Knowing that a string with zero width is a mathematical idealization, we introduce
a Gaussian-type smearing function \( W(q) \), normalized as \( \int W(q)4\pi q^{2}dq=1 \).
So we define 
\begin{equation}
f(\vec{q},\vec{p})=\delta (\vec{p}-\overrightarrow{\Delta P})W\left( \left\Vert \vec{q}-\vec{Q}\right\Vert \right) ,
\end{equation}
with the norm in hyperbolic coordinates being given as
\begin{equation}
\left\Vert \vec{q}\right\Vert =\left( q_{1}\right) ^{2}+\left( q_{2}\right) ^{2}+\tau ^{2}\left( q_{3}\right) ^{2}.
\end{equation}
The energy momentum tensor for the complete string is then just the sum over
all string pieces, which gives
\begin{equation}
T^{\mu \nu }(\vec{q})=\sum _{\mathrm{string}\, \mathrm{segments}}\frac{\Delta P^{\mu }\Delta P^{\nu }}{P^{0}}\, W\left( \left\Vert \vec{q}-\vec{Q}\right\Vert \right) .
\end{equation}
 We define a local comoving frame via the four-velocity 
\begin{equation}
u^{\mu }(\vec{q})=\frac{n^{\mu }(\vec{q})}{\sqrt{n^{\mu }(\vec{q})\, n_{\mu }(\vec{q})}}
\end{equation}
 with
\begin{equation}
n^{\mu }(\vec{q})=T^{0\mu }(\vec{q}).
\end{equation}
 This allows then the calculation of the energy density in the comoving frame,
\begin{equation}
\epsilon (q)=T^{\mu \nu }(\vec{q})\, u_{\mu }(\vec{q})\, u_{\nu }(\vec{q}),
\end{equation}
 and the flow velocity
\begin{equation}
\vec{v}(q)=\vec{u}(q)/u^{0}.
\end{equation}

\section{Results}

In the following, we show results for different high energy reactions. All the
calculations are done at the proper time \( \tau = \) 1fm/c, and we suppress
writing this variable in the following. We use always a width \( \sigma =0.5\textrm{fm} \)
for smearing function \( W(q) \). We define the rapidity field \( y_{z}(x,y,\eta ) \)
as
\begin{equation}
y_{z}(x,y,\eta )=\frac{1}{2}\log \frac{1+v_{z}(x,y,\eta )}{1-v_{z}(x,y,\eta )},
\end{equation}
where \( v_{z}(x,y,v_{z}) \) is the \( z \) component of the velocity field. 

\begin{figure}[htb]
{\par\centering \resizebox*{!}{0.26\textheight}{\includegraphics{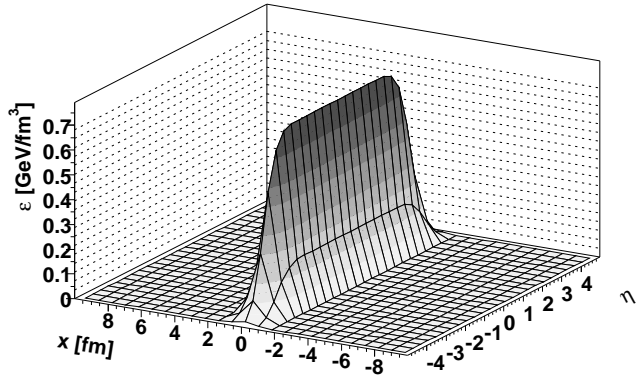}} \par}

\caption{Energy density in the \protect\( x-\eta \protect \) plane at \protect\( y=0\protect \)
for a single string of energy 50 GeV. \label{string-ene-xeta} }
\end{figure}

\begin{figure}[htb]
{\par\centering \resizebox*{!}{0.26\textheight}{\includegraphics{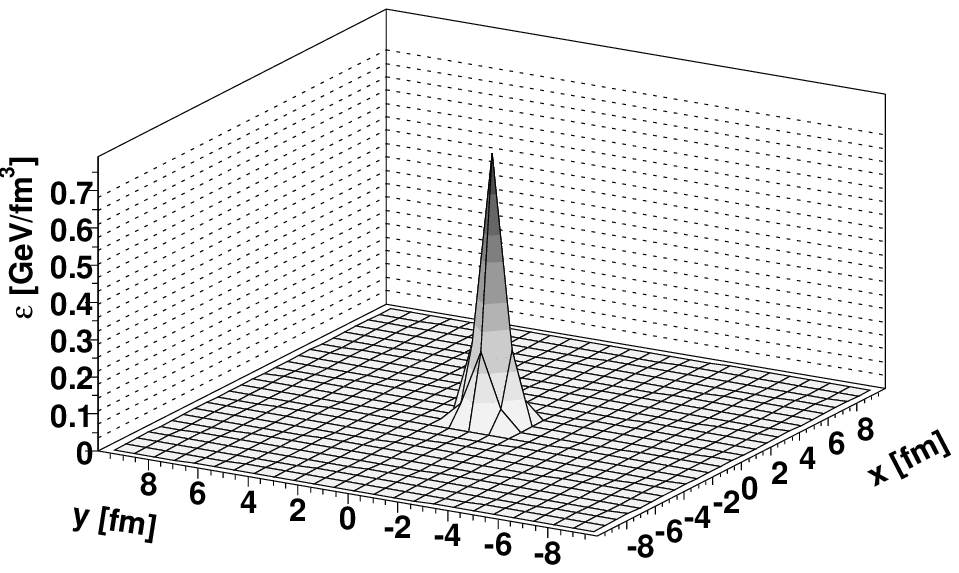}} \par}

\caption{Energy density in the \protect\( x-y\protect \) plane at \protect\( \eta =0\protect \)
for a single string of energy 50 GeV. \label{string-ene-xy}}
\end{figure}

\begin{figure}[htb]
{\par\centering \resizebox*{!}{0.26\textheight}{\includegraphics{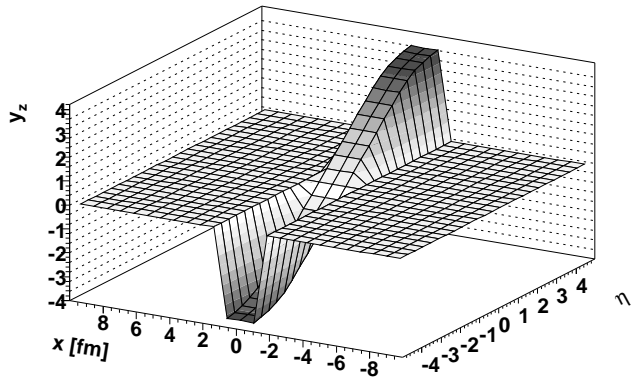}} \par}

\caption{Rapidity field in the \protect\( x-\eta \protect \) plane at \protect\( y=0\protect \)
for a single string of energy 50 GeV. \label{string-flo-xeta}}
\end{figure}

As a reference, we first show results for a single string without kinks, having
an energy of 50 GeV, as it may occur in electron-positron annihilation. In fig.
\ref{string-ene-xeta}, we plot the energy density \( \epsilon (x,y,\eta ) \)
as a function of \( x \) and \( \eta  \) for \( y=0 \). As expected, the
energy density is peaked around \( x=0 \), whereas it is distributed evenly
in \( \eta  \) between limits defined by the energy of the string. Since there
is a cylindrical symmetry with respect to the \( z \) axis, it is no surprise
that the energy density as a function of \( x \) and \( y \) for \( \eta =0 \)
shows a narrow peak around \( x=y=0 \), as seen in fig. \ref{string-ene-xy}.
The rapidity field \( y_{z}(x,y,\eta ) \) as a function of \( x \) and \( \eta  \)
for \( y=0 \) is shown in fig. \ref{string-flo-xeta}, where we observe roughly
\( y_{z}(x,y,\eta )=\eta  \). The other components of the velocity field are
zero. 

Kinky strings show a very similar behavior. Due to the transverse momenta introduced
via the kinks, the cylindrical symmetry is slightly distorted, and we observe
as well very small but finite values for the transverse components of the velocity
fields.Since the results are so close to the simple string discussed above,
we do not show the figures here.

Next we consider proton-proton scattering at 50 GeV. 
\begin{figure}[htb]
{\par\centering \resizebox*{!}{0.26\textheight}{\includegraphics{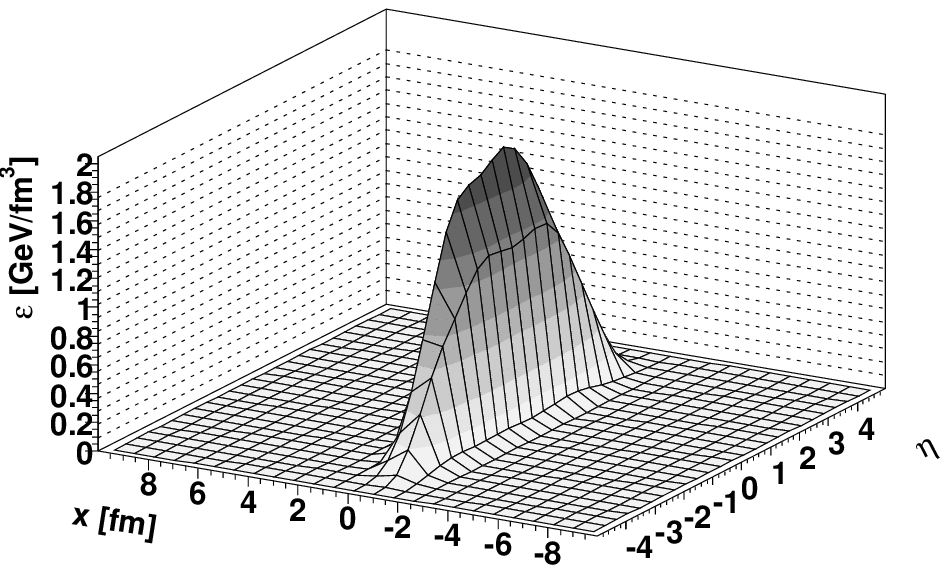}} \par}

\caption{Energy density in the \protect\( x-\eta \protect \) plane at \protect\( y=0\protect \)
for a proton-proton collision at 50 GeV. \label{pp-ene-xeta}}
\end{figure}
\begin{figure}[htb]
{\par\centering \resizebox*{!}{0.26\textheight}{\includegraphics{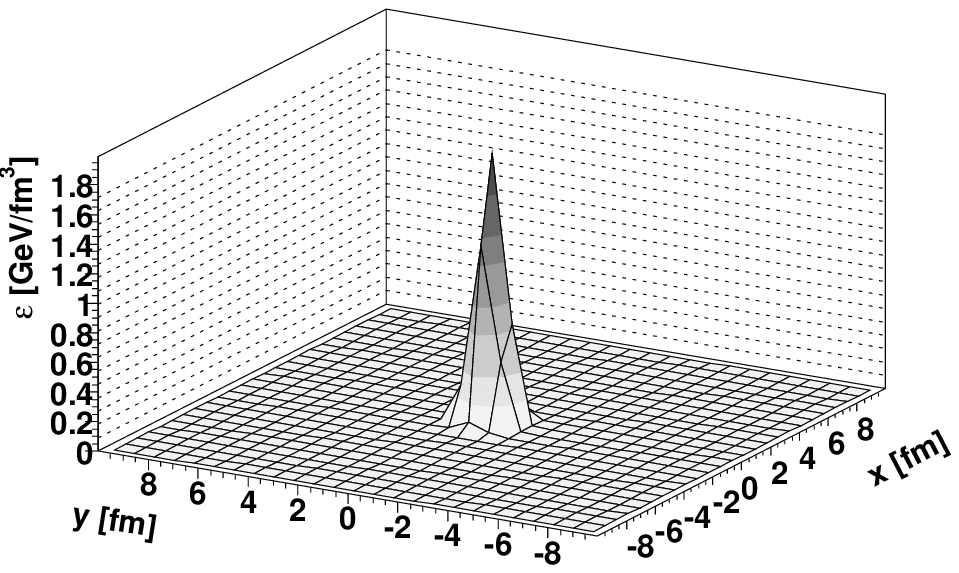}} \par}

\caption{Energy density in the \protect\( x-y\protect \) plane at \protect\( \eta =0\protect \)
for a proton-proton collision at 50 GeV. \label{pp-ene-xy}}
\end{figure}
\begin{figure}[htb]
{\par\centering \resizebox*{!}{0.26\textheight}{\includegraphics{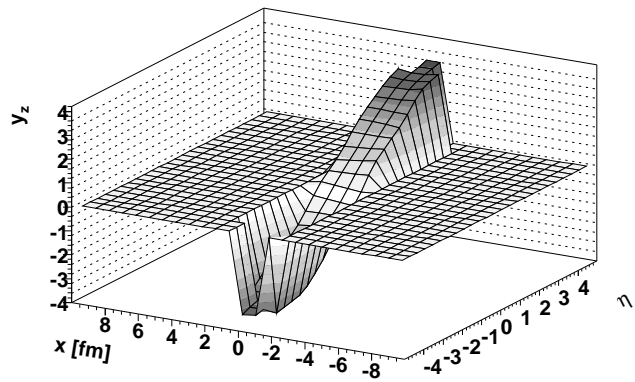}} \par}

\caption{Rapidity field in the \protect\( x-\eta \protect \) plane at \protect\( y=0\protect \)
for a proton-proton collision at 50 GeV.\label{pp-flo-xeta}}
\end{figure}
Here we have usually several strings, the minimum number being two, plus two
remnants (which may be strings in case of high mass excitation). In the example
considered here, we have altogether four strings, with respective energies of
17.2 GeV, 9.3 GeV, 7.6 GeV and 15.8 GeV. Since the strings have different masses
and are not sitting in the cms system, the energy density \( \epsilon (x,y,\eta ) \)
as a function of \( x \) and \( \eta  \) for \( y=0 \) does not show such
a flat behavior as in the case of a pure string, as seen in fig. \ref{pp-ene-xeta},
but we observe a peak at \( \eta =0 \), whose value is roughly 2.5 times bigger
than the one for a string. However, the energy density as a function of \( x \)
and \( y \) for \( \eta =0 \) shows as well a narrow peak around \( x=y=0 \),
as shown in fig. \ref{pp-ene-xy}, and the rapidity field \( y_{z}(x,y,\eta ) \)
is roughly equal to \( \eta  \) , see fig. \ref{pp-flo-xeta}. 

\begin{figure}[htb]
{\par\centering \resizebox*{!}{0.26\textheight}{\includegraphics{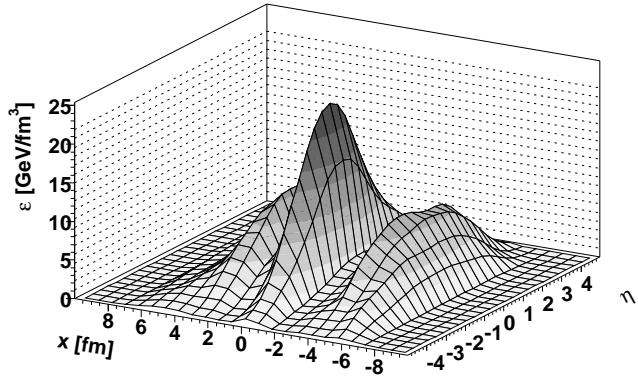}} \par}

\caption{Energy density in the \protect\( x-\eta \protect \) plane at \protect\( y=0\protect \)
for a central gold-gold collision at 200 AGeV. \label{auau-ene-xeta}}
\end{figure}

\begin{figure}[htb]
{\par\centering \resizebox*{!}{0.26\textheight}{\includegraphics{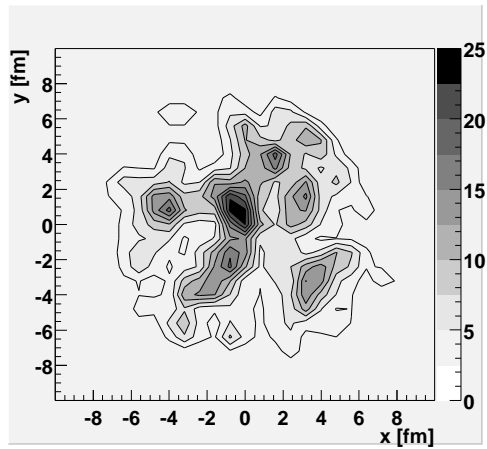}} \par}

\caption{Energy density in the \protect\( x-y\protect \) plane at \protect\( \eta =0\protect \)
for a central gold-gold collision at 200 AGeV. \label{auau-ene-xy}}
\end{figure}

\begin{figure}[htb]
{\par\centering \resizebox*{!}{0.26\textheight}{\includegraphics{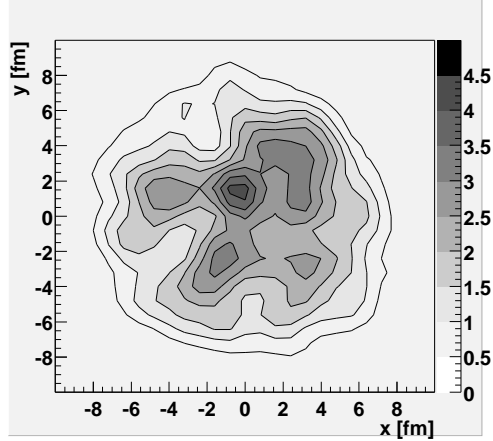}} \par}

\caption{Projected nucleon density in the \protect\( x-y\protect \) plane for a central
gold-gold collision at 200 AGeV. \label{auau-nuc-xy}}
\end{figure}

Let us consider gold-gold collisions at RHIC (200 AGeV). In fig. \ref{auau-ene-xeta}, 
we show the energy density \( \epsilon (x,y,\eta ) \) as a function of \( x \)
and \( \eta  \) for \( y=0 \). For a fixed value of \( x \), we observe a
similar shape as for proton-proton scattering: a broad distribution with a smooth
peak around zero. Of course, the magnitude is much bigger. Considering, however,
the variation with \( x \) for given \( \eta  \), we observed large fluctuations:
pronounced peaks followed by deep valleys. If we regard the energy density as
a function of \( x \) and \( y \) for \( \eta =0 \), as shown in fig. \ref{auau-ene-xy}, 
we observe correspondingly several peaks overlaying the general roughly rotationally
symmetric structure, which increases towards the center. To investigate the
origin of these fluctuations, we show in fig. \ref{auau-nuc-xy} the distribution
of the number of 
nucleons (projectile plus target) projected to the plane \( z=0 \), in units
nucleons per \( \mathrm{fm}^{2} \). From a spherically symmetric nuclear density,
we expect a rotationally symmetric distribution of this projection, increasing
towards the center (\( x=y=0 \)). This is also what one observes, roughly.
But looking more closely, we clearly observe large fluctuations with pronounced
peaks. And even more, these peaks correspond exactly to the peaks in the energy
density distribution, which proves that the fluctuations in energy density are
due to geometrical fluctuations in the distribution of nucleons. 
\begin{figure}[htb]
{\par\centering \resizebox*{!}{0.26\textheight}{\includegraphics{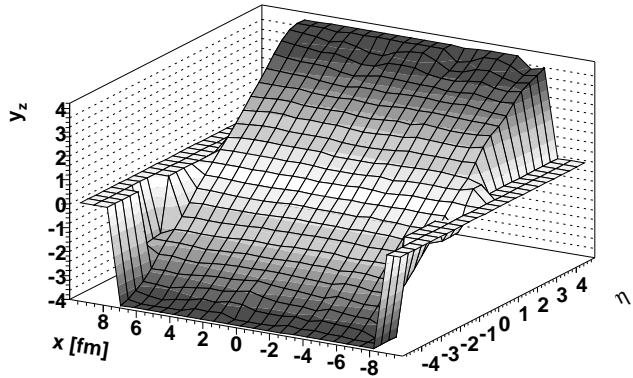}} \par}

\caption{Rapidity field in the \protect\( x-\eta \protect \) plane at \protect\( y=0\protect \)
for a central gold-gold collision at 200 AGeV. \label{auau-flo-xeta}}
\end{figure}
\begin{figure}[htb]
{\par\centering \resizebox*{!}{0.26\textheight}{\includegraphics{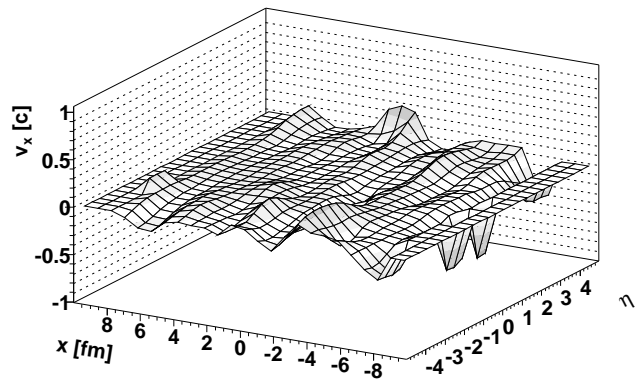}} \par}

\caption{Velocity field (\protect\( x\protect \)-component) in the \protect\( x-\eta \protect \)
plane at \protect\( y=0\protect \) for a central gold-gold collision at 200
AGeV. \label{auau-flox-xeta}}
\end{figure}
We finally consider rapidity and velocity fields: in fig. \ref{auau-flo-xeta}
the rapidity field \( y_{z}(x,y,\eta ) \) and in fig. \ref{auau-flox-xeta}
the \( x \) component \( v_{x}(x,y,\eta ) \) of the velocity field, both as
a function of \( x \) and \( \eta  \) for \( y=0 \). We observe roughly \( y_{z}=\eta  \)
and \( v_{x} \) close to zero. So the velocity field is practically purely
longitudinal and shows the so-called ``Bjorken scaling'' (\( y_{z}=\eta  \)).

\section{Summary}

We presented a new way to calculate macroscopic quantities like energy density
or velocity (or rapidity) fields at an early stage of a nucleus-nucleus collision
at ultra-relativistic energies. The calculation is based on a sophisticated
treatment of the primary interactions, when the two nuclei are traversing each
other, using the parton-based Gribov-Regge model \textsc{neXus}. The important
point is an appropriate treatment of soft partons, which contribute substantially
and which are usually completely neglected. In \textsc{neXus}, soft and hard
physics are considered consistently: hard partons are treated explicitly based
on pQCD, soft ones are included implicitly, using the string picture. This allows
a quite reliable calculation of the above-mentioned macroscopic quantities.

We analyzed single strings, proton-proton scattering and heavy ion collisions.
In the latter case, we find almost perfect ``Bjorken scaling'': the rapidity
\( y_{z} \) coincides with the space-time rapidity \( \eta  \), whereas the
transverse flow is practically zero. This is often employed as initial condition
for hydrodynamical treatments. However, the \( \eta  \) dependence of the energy
density does not show a well defined plateau corresponding to ``boost invariance''.
Furthermore, the distribution of the energy density in the transverse plane
shows typically a very ``bumpy'' structure, which fluctuates considerably
from event to event. 

This work has been funded in part by the IN2P3/CNRS (PICS 580) and the Russian
Foundation of Basic Researches (RFBR-98-02-22024). H.J.D. acknowledges support
from NASA grant number NAG-9246. 

\bibliographystyle{pr2}
\bibliography{a}

\end{document}